\documentclass[journal]{IEEEtran}

\usepackage{amssymb}
\usepackage{amsmath}
\usepackage{cite}
\usepackage{graphicx}
\usepackage{tabularx,booktabs,multirow,array}
\usepackage[caption=false,font=footnotesize]{subfig}
\usepackage{hyperref}
\DeclareSubrefFormat{parens}{#1(#2)}
\usepackage{bm}

\hypersetup{
  hidelinks,
  pdftitle={Security-Constrained Operation of IBR-Dominated Power Systems: Static and Dynamic Security Across Preventive and Corrective Decisions},
  pdfauthor={Buxin She},
  pdfkeywords={corrective control, dynamic security, inverter-based resources, preventive scheduling, security-constrained scheduling, system strength}
}

\newcolumntype{Y}{>{\raggedright\arraybackslash}X}
\newcolumntype{C}[1]{>{\centering\arraybackslash}p{#1}}
\newcolumntype{L}[1]{>{\raggedright\arraybackslash}p{#1}}

\begin{document}

\title{Security-Constrained Operation of IBR-Dominated Power Systems:
Static and Dynamic Security Across Preventive and Corrective Decisions}

\author{Buxin~She,~\IEEEmembership{Member,~IEEE}%
\thanks {This work was developed within the IEEE Task Force on Virtual Inertia Scheduling and Control (VISC) for IBR-Dominant Power Systems.}
}

\markboth{Preprint}%
{She: Security-Constrained Operation in IBR-Dominated Power Systems}

\maketitle

\begin{abstract}
Inverter-based resources (IBRs) couple power system operation to fast dynamics and controller-dependent responses.
Their configurable capabilities are reshaping the formulation and coordination of security-constrained operation.
This paper presents a two-axis view: static versus dynamic security and preventive versus corrective decision timing.
Preventive scheduling is extending from static post-contingency feasibility toward dynamic security, while corrective operation spans equilibrium-based corrective actions and trajectory-based control.
A generic formulation represents this change and makes the preventive--corrective tradeoff explicit.
Existing formulations, methods, and capability representations are reviewed and synthesized within this framework.
The surveyed work yields two findings.
First, IBR capability can expand the feasible set or relieve security constraints, reducing operating cost or improving security performance.
Second, shared capability and constraints across formulations determine whether that capability remains operationally deliverable.
These findings motivate future research in IBR capability characterization and quantification, joint scheduling, and scalable solution frameworks.
\end{abstract}


\IEEEpeerreviewmaketitle

\section*{Nomenclature}
\begin{IEEEdescription}[\setlength{\labelsep}{1em}\IEEEsetlabelwidth{$\boldsymbol\Phi_c$}]
\setlength{\itemsep}{0pt}
\setlength{\parsep}{0pt}
\item[$c$] Contingency index.
\item[$\mathcal C$] Set of prescribed contingencies.
\item[$\boldsymbol{\vartheta}$] Preventive schedule and resource-configuration vector.
\item[$\pi_c$] Corrective action for contingency $c$.
\item[$C_0$] Pre-contingency operating cost.
\item[$C_c$] Cost or consequence associated with contingency $c$.
\item[$\rho$] Balance weight of preventive--corrective cost.
\item[$\mathcal F_0$] Pre-contingency feasible set.
\item[$\mathcal S_c$] Security set for contingency $c$.
\item[$\mathcal A_c$] Feasible corrective-action set for contingency $c$.
\item[$\mathbf z_c$] Vector of post-contingency differential and algebraic variables.
\item[$\mathbf a_c$] Static corrective-action vector for contingency $c$.
\item[$\boldsymbol\mu_c$] Preconfigured automatic-control law for contingency $c$.
\item[$\Gamma_c$] Function that determines the post-contingency initial condition from $\boldsymbol{\vartheta}$.
\item[$\mathbf F_c$] Function defining the differential-algebraic network and equipment equations for contingency $c$.
\item[$\boldsymbol\Phi_c$] Dynamic-security constraint vector for contingency $c$.
\end{IEEEdescription}

\newpage

\section{Introduction}
\label{sec:intro}

\IEEEPARstart{P}{ower} system security depends on coordinated actions before and after a prescribed contingency.
North American Electric Reliability Corporation (NERC) specifies an Operational Planning Analysis and associated operating plans for anticipated conditions~\cite{NERCTOP002}.
Within this process, preventive scheduling denotes the optimization-based selection of pre-contingency commitment, dispatch, topology, and operating configuration.
After a disturbance, preconfigured automatic response and real-time assessment precede post-contingency corrective action~\cite{li2016real}.

Classical security-constrained operation represents much of this chain through static post-contingency feasibility.
Security-constrained unit commitment (SCUC), security-constrained economic dispatch (SCED), and security-constrained optimal power flow (SCOPF) impose network and operating limits for prescribed contingencies~\cite{Alsac1974OPF, Stott1987Security, Monticelli1987PCSCOPF}.
DC formulations enforce active-power balance, line-flow limits, and applicable reserve or ramping requirements.
AC formulations additionally represent reactive-power balance and voltage constraints.
Preventive formulations retain a common pre-contingency schedule, whereas corrective SCOPF includes contingency-dependent redispatch or network actions.
Both assess an acceptable post-contingency state under the adopted network model, but not necessarily a secure trajectory~\cite{Wood2013, Chen2023SCUC, Latify2025SCUC}.

\begin{figure*}[t]
\centering
\includegraphics{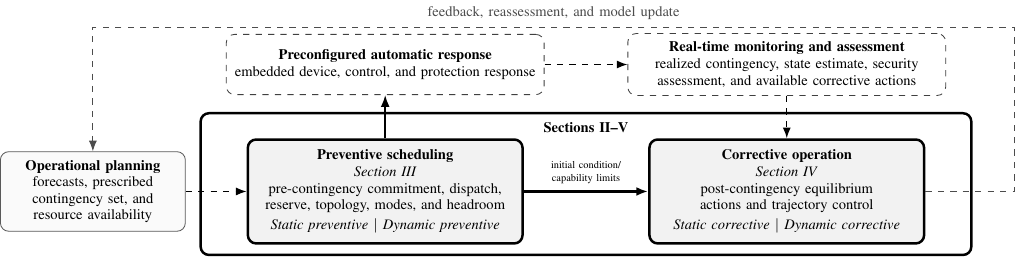}
\caption{Operational framework of IBR-dominated power systems. Solid arrows denote decision effects; dashed arrows denote information and feedback.}
\label{fig:ibr_operation_framework}
\end{figure*}

Security-constrained operation is therefore going beyond static post-contingency feasibility.
Dynamic security constraints are increasingly embedded in preventive scheduling~\cite{She2024VISRTED, She2025VISMicrogrid, Wang2025DynamicsScheduling}, while corrective operation also includes trajectory-based control~\cite{Genc2010PreventiveCorrective,Martin2017CorrectiveMPC}.
The large-scale deployment of IBRs changes system dynamics and expands the available decision space.
These resources include wind, solar photovoltaic generation, and battery energy storage systems (BESSs), together with other converter-interfaced assets such as high-voltage direct current (HVDC) links~\cite{Milano2018LowInertia}.
IBR operating points, grid-following (GFL) and grid-forming (GFM) modes, and control parameters can be adjusted before or after a contingency~\cite{she2023inverter}.
These adjustments determine the dynamic support capabilities available at each decision stage~\cite{Lin2020GFMRoadmap, Rosso2021GFMReview}.
Preventive decisions configure these capabilities and corrective decisions may update setpoints or control policies after a contingency.
These developments increasingly couple operating functions that were often modeled separately.

Existing reviews generally address one of three areas: SCUC~\cite{Chen2023SCUC,Latify2025SCUC}, SCOPF and AC-OPF~\cite{Capitanescu2011SCOPF,Capitanescu2016ACOPF,Aravena2023GO}, or stability and online security assessment~\cite{Morison2004SecurityAssessment,Hatziargyriou2021Stability,DeCaro2023DataDrivenSecurity}.
IBR-oriented surveys mainly examine GFM controls and stability mechanisms~\cite{Rosso2021GFMReview}, while reviews of dynamic-security-constrained optimization focus largely on transient-stability-constrained OPF~\cite{Zhang2025TSCOPF}.
The literature does not compare security representation and decision timing within a common framework.
It also does not explain how scheduled IBR capability enables and constrains corrective action.

To address this gap, this paper classifies preventive and corrective decisions under static and dynamic security.
Operational planning supplies the forecasts, prescribed contingency set, and resource-availability information used by preventive scheduling.
Preconfigured automatic response changes the post-contingency state.
Real-time assessment identifies that state and the corrective actions that remain available.
Fig.~\ref{fig:ibr_operation_framework} shows how system resources, particularly IBR capabilities, are configured before a contingency, deployed after the event, and coordinated with corrective action.
The solid box marks the focus of this review, including preventive scheduling, corrective operation, and their coupling.
The contributions of this paper are as follows.
\begin{itemize}
    \item A two-axis taxonomy classifies security-constrained operation by security representation and decision timing.
    \item A generic formulation links the preventive solution to the post-contingency initial condition, and feasible corrective decisions under shared capability limits.
    \item The literature is reviewed and compared in terms of their security criteria, decisions, methods, and solution frameworks.
    \item A synthesis distinguishes IBR capability effects within individual formulations from joint effects across coupled formulations.
\end{itemize}

The remainder of this paper is organized as follows.
Section~\ref{sec:two_axis} presents the two-axis framework and generic formulation. Sections~\ref{sec:preventive} and \ref{sec:corrective} examine preventive and corrective operation, respectively. Section~\ref{sec:closed_loop} discusses the IBR capability effects and research directions, and Section~\ref{sec:conclusion} concludes the paper.
\section{Two-Axis Taxonomy and Generic Formulation of Security-Constrained Operation}
\label{sec:two_axis}

This section introduces the two-axis taxonomy and a generic formulation that represents its four quadrants within a common preventive--corrective structure.

\subsection{Taxonomy and Trends}

\begin{figure*}[t]
\centering
\includegraphics{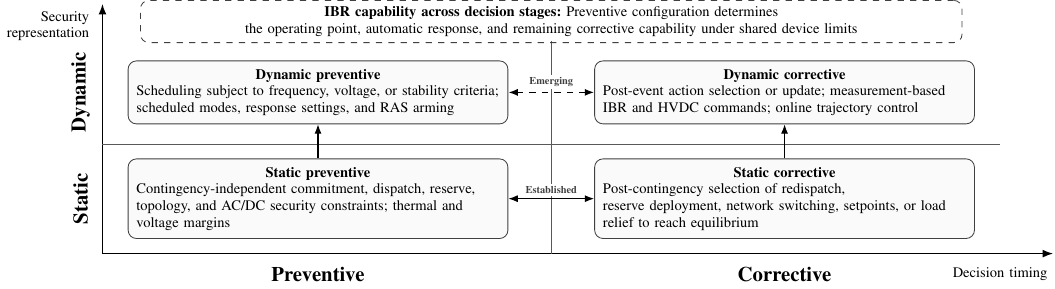}
\caption{Two-axis taxonomy of security-constrained operation.}
\label{fig:quadrants}
\end{figure*}

\subsubsection{Two-Axis Taxonomy}
As shown in Fig.~\ref{fig:quadrants}, the two axes separate security representation from decision timing.
Static security denotes steady-state constraints and equilibrium feasibility, whereas dynamic security denotes post-disturbance metrics, stability margins, or trajectories~\cite{Kundur2004Definition,Hatziargyriou2021Stability,NERC2025EMT}.
The classification follows the security property being assessed.
An analytically derived operating boundary, surrogate-model boundary, or algebraic constraint remains a dynamic-security representation when it is validated to assess a dynamic property.

The preventive--corrective axis classifies the information available when a decision is made.
A preventive decision is fixed before a contingency, whereas a corrective decision uses the identified contingency or post-contingency measurements~\cite{Wehenkel2004PreventiveEmergency,Genc2010PreventiveCorrective}.
A post-event policy may be designed offline, but its realized action is selected after the event.
Controller settings, remedial action scheme (RAS) logic, and RAS arming fixed before the event are preventive decisions.
Their post-event execution remains automatic.
SCUC, SCED, and SCOPF denote formulation families.
Their decisions are preventive when they are common across contingencies and corrective when they are selected or updated after the contingency is identified.
NERC terminology describes a RAS as automatically taking corrective actions~\cite{NERC2020RAS}.
This paper uses \emph{corrective action} more narrowly for a decision selected using post-event information.

\subsubsection{Emerging Trends}
Fig.~\ref{fig:quadrants} also locates representative methods in the four quadrants, which are examined in Sections~\ref{sec:preventive} and \ref{sec:corrective}.
Two trends emerge, both driven by IBR flexibility and controllability.
First, security-constrained operation is extending from established static formulations toward dynamic-security-constrained scheduling and control.
The operating point, control mode, and verified response of IBRs can be scheduled and jointly determine steady-state margins and dynamic response.
This static--dynamic coupling motivates explicit dynamic security constraints~\cite{Badesa2019FrequencyServices, Tuo2023Locational, Wang2023MultiAreaFCUC, Liu2024Nadir}.
Second, preventive scheduling and corrective control are becoming more tightly coupled.
IBR capability can be configured before a contingency, while corrective setpoints and control parameters can be updated using post-contingency information.
Consequently, preventive decisions determine the feasible corrective action set, while anticipated corrective capability feeds back into the preventive schedule~\cite{Matevosyan2019GFM, IEEE2800_2022, AEMO2025SecurityEnablement, Cui2025GFM}.

\subsection{Generic Mathematical Formulation}
\label{sec:math}

Consider the following generic preventive--corrective formulation.
The preventive vector $\boldsymbol{\vartheta}$ is common to all contingencies, while $\pi_c$ denotes the corrective action.

\subsubsection{Formulation}

A compact form is
\begin{align}
\min_{\boldsymbol{\vartheta},\{\pi_c\}_{c\in\mathcal C}}\quad
& C_0(\boldsymbol{\vartheta})+
\rho\!\left(\{C_c(\boldsymbol{\vartheta},\pi_c)\}_{c\in\mathcal C}\right) \label{eq:unified_obj}\\
\text{s.t.}\quad
& \boldsymbol{\vartheta}\in\mathcal F_0, \label{eq:pre_constraints}\\
& (\boldsymbol{\vartheta},\pi_c)\in\mathcal S_c,\quad
\pi_c\in\mathcal A_c(\boldsymbol{\vartheta}),
\quad c\in\mathcal C. \label{eq:post_constraints}
\end{align}
The objective combines the pre-contingency cost $C_0$ with the corrective costs $C_c$ for contingency $c$.
The sets $\mathcal F_0$, $\mathcal S_c$, and $\mathcal A_c$ impose pre-contingency feasibility, contingency security, and corrective-decision feasibility, respectively.
Eqs.~\eqref{eq:unified_obj}--\eqref{eq:post_constraints} form a deterministic equivalent of the classical formulation~\cite{Monticelli1987PCSCOPF,Capitanescu2008Corrective,Capitanescu2011SCOPF}.
The corrective decision $\pi_c$ may be designed offline but is realized only after contingency $c$ is identified, using the information allowed by $\mathcal A_c$.

\subsubsection{Static and Dynamic Security Modules}

The set $\mathcal S_c$ can use a static or dynamic security representation.
For static security, $\pi_c$ is a corrective-action vector $\mathbf a_c$, and
\begin{equation}
\begin{aligned}
\mathcal S_c^{\mathrm{stat}}
=\bigl\{(\boldsymbol{\vartheta},\mathbf a_c)\ \big|\ &
\exists\,\mathbf x_c:\
\mathbf g_c(\mathbf x_c;\boldsymbol{\vartheta},\mathbf a_c)=\mathbf 0,\\
&\mathbf h_c(\mathbf x_c;\boldsymbol{\vartheta},\mathbf a_c)\leq\mathbf 0
\bigr\}.
\end{aligned}
\label{eq:static_module}
\end{equation}
Here, $\mathbf x_c$ is the post-contingency equilibrium state, and $\mathbf g_c$ and $\mathbf h_c$ impose the balance, network, and equipment constraints.
The set $\mathcal A_c(\boldsymbol{\vartheta})$ limits the corrective action according to the scheduled condition and the preconfigured response represented at equilibrium.
This representation can include the equilibrium effect of a preconfigured response but does not ensure its trajectory.

For dynamic security, $\mathbf z_c$ collects the differential and algebraic variables.
\begin{equation}
\begin{aligned}
\mathbf z_c(0^+)
&=\Gamma_c(\boldsymbol{\vartheta}),\\
\mathbf 0
&=\mathbf F_c\!\bigl(t,\dot{\mathbf z}_c(t),\mathbf z_c(t);
\boldsymbol{\vartheta},\boldsymbol\mu_c,\pi_c\bigr),
\quad t\in[0,T_c].
\end{aligned}
\label{eq:dynamic_module}
\end{equation}
The function $\Gamma_c$ sets the post-contingency initial condition, and $\mathbf F_c=\mathbf0$ represents the network and equipment dynamics over the assessment period $T_c$.
The preventive configuration fixes $\boldsymbol\mu_c$, whose automatic response can vary with the measured state.
The policy $\pi_c$ instead selects an action from post-contingency information.
A feedback law fixed before the event remains part of $\boldsymbol\mu_c$.
Selected dynamic-security constraints are written compactly as
\begin{equation}
\boldsymbol\Phi_c\!\left(
\boldsymbol{\vartheta},\mathbf z_c(\cdot)
\right)\leq\mathbf 0.
\label{eq:dynamic_metric}
\end{equation}
The vector $\boldsymbol\Phi_c$ imposes path or terminal stability and response requirements~\cite{Gan2000TSCOPF,Capitanescu2009DynamicSimulation}.
The set $\mathcal S_c^{\mathrm{dyn}}$ contains pairs that satisfy \eqref{eq:dynamic_module} and \eqref{eq:dynamic_metric}.
The set $\mathcal A_c(\boldsymbol{\vartheta})$ specifies the available information and response time, together with the device capability remaining after automatic response.

\subsubsection{Four-Quadrant Representation and Coupling}

Static preventive and corrective operation use $\mathcal S_c^{\mathrm{stat}}$ with $\mathbf a_c=\mathbf0$ and $\mathbf a_c\in\mathcal A_c(\boldsymbol{\vartheta})$, respectively.
Dynamic preventive and corrective operation use $\mathcal S_c^{\mathrm{dyn}}$ with $\pi_c\equiv0$ and $\pi_c\in\mathcal A_c(\boldsymbol{\vartheta})$, respectively.
In both preventive cases, the preconfigured automatic response remains active.
The vector $\boldsymbol{\vartheta}$ determines the base schedule, post-contingency initial condition, automatic response, and feasible corrective decisions.
Accordingly, $\mathcal S_c$ imposes the modeled security requirements, while $\mathcal A_c$ limits the corrective decision used to satisfy them.

\begin{table*}[t]
\caption{Literature summary of static preventive operation}
\label{tab:static_formulations}
\centering
\scriptsize
\begin{tabularx}{\textwidth}{L{0.15\textwidth} L{0.17\textwidth} L{0.19\textwidth} L{0.21\textwidth} Y}
\toprule
\textbf{Topics} & \textbf{Security criteria} & \textbf{Scheduled decisions} & \textbf{Methods and literature} & \textbf{Strengths/Limitations} \\
\midrule
Preventive DC-SCUC/SCED & Linear base-case and $N$-$1$ flows; reserve/ramp feasibility & Commitment, dispatch, reserve, and topology & MILP-based SCUC/SCED~\cite{Stott1987Security,Ostrowski2012TightUC,MoralesEspana2013TightUC,Chen2023SCUC,Latify2025SCUC} & Scalable, transparent; AC voltage/reactive power and dynamics not represented \\
Preventive AC-SCOPF/AC-SCUC & AC equilibria; voltage/reactive limits; loadability margins & Active/reactive dispatch, voltage controls, and commitment & Nonlinear/mixed-integer AC models~\cite{Alsac1974OPF,Fu2005ACSCUC,Capitanescu2011SCOPF,Capitanescu2016ACOPF,VanCutsem1998Voltage,Chiang1995Voltage,Aravena2023GO,Curtis2023SCACOPF} & Higher steady-state fidelity; nonconvexity and contingency scale \\
Stochastic security scheduling & Renewable/load scenarios; contingency feasibility & Commitment and reserve; prescribed balancing rule for forecast uncertainty & Scenario-based scheduling~\cite{Wang2008WindSCUC,Morales2009Reserves,Papavasiliou2011Reserve,Bouffard2008StochasticSecurity,Morales2014Renewables} & Forecast uncertainty; scenario growth and stage interpretation \\
Chance-constrained and risk-based operation & Probabilistic line/voltage/reserve limits; severity weighting & Dispatch, reserve, participation factors, storage state & Chance/risk formulations~\cite{Roald2015Weighted,Lubin2016ChanceACOPF} & Reliability--risk tradeoff; tail modeling and calibration \\
Adaptive robust SCUC & Worst-case uncertainty; adaptive balancing feasibility & Commitment and reserve; prescribed balancing rule & Robust/decomposition methods~\cite{Jiang2012RobustUC,Bertsimas2013RobustSCUC} & Limited distribution assumptions; conservatism and balancing feasibility \\
\bottomrule
\end{tabularx}
\end{table*}
\section{Security-Constrained Preventive Operation}
\label{sec:preventive}

This section examines static and dynamic preventive operation, followed by solution frameworks for embedding the corresponding security constraints in scheduling.

\subsection{Static Preventive Operation}

Static preventive formulations determine pre-contingency topology, commitment, and other scheduled controls such that every prescribed contingency has an acceptable equilibrium.
The scheduled decisions are common across contingency scenarios.
Table~\ref{tab:static_formulations} summarizes the classical formulations and their uncertainty extensions.

\subsubsection{SCUC, SCED, and SCOPF With Common Preventive Decisions}

SCUC determines unit commitment and generation schedules over multiple periods.
Its constraints typically include power balance, equipment ratings, minimum up- and down-times, ramping, reserve, and network security~\cite{Ostrowski2012TightUC, MoralesEspana2013TightUC}.
SCED updates dispatch over shorter horizons, typically with commitment fixed~\cite{Yang2023Dispatch}.
SCOPF emphasizes network feasibility, including active and reactive power balance, voltage limits, and contingency constraints~\cite{Capitanescu2011SCOPF, Capitanescu2016ACOPF}.
AC-SCUC combines the multiperiod integer decisions of SCUC with AC network and contingency constraints, whereas AC-SCOPF focuses primarily on secure AC operating points~\cite{Fu2005ACSCUC,Capitanescu2016ACOPF}.
These formulations differ mainly in scheduling horizon, integer decisions, and network fidelity.
For IBRs, the schedule sets the $P$--$Q$ operating point, reserve, or enabled control mode~\cite{Wen2016StorageSCUC,She2024VISRTED,She2025VISMicrogrid,Cui2025ControlMode}.
Power and energy limits link these variables to the response available after a contingency.
Static models represent the resulting equilibrium tradeoff but not the dynamic performance.

Within this static representation, voltage security is enforced through bus voltage and reactive power limits at the prescribed post-contingency equilibria.
Voltage-stability-constrained OPF may further impose $P$--$V$ or $Q$--$V$ margins, singularity measures, or continuation-power-flow limits~\cite{VanCutsem1998Voltage,Chiang1995Voltage}.
These measures can be approximated by surrogate constraints.

Large-scale scheduling commonly uses the DC approximation with power-transfer distribution factors (PTDFs) and line-outage distribution factors (LODFs)~\cite{Stott1987Security}.
It yields a mixed-integer linear program (MILP) for SCUC and a linear or quadratic program for SCED.
AC security constraints can be enforced directly or through decomposition~\cite{Fu2005ACSCUC, Capitanescu2016ACOPF}.
Convex relaxations and approximations reduce the computational burden~\cite{Low2014ConvexPartI,Coffrin2016QC,Molzahn2019ConvexSurvey}.
Post-solution validation with iterative refinement provides another practical option~\cite{Aravena2023GO}.
The Advanced Research Projects Agency--Energy (ARPA-E) Grid Optimization (GO) Competition demonstrated large-scale solution of realistic security-constrained AC optimal power flow (OPF)~\cite{Aravena2023GO,Curtis2023SCACOPF}.
Common solution methods use contingency screening~\cite{Capitanescu2007Filtering,Ardakani2013Umbrella}, decomposition and parallel computation~\cite{Velloso2021Decomposition,Gholami2023ADMM}, or asynchronous surrogate-model assessment~\cite{Petra2023Surrogate}.
These methods scale well for DC models, but nonconvex AC security constraints remain computationally difficult.

\subsubsection{Uncertainty Extensions}

Static preventive scheduling faces uncertainty in forecasts, resource availability, and contingencies.
For IBRs, capability uncertainty also reflects available energy and equipment status.
Scheduling models represent uncertainty through scenarios, uncertainty sets, chance constraints, or risk measures.

Stochastic scheduling represents renewable and load uncertainty through scenarios and probabilities~\cite{Bouffard2008StochasticSecurity, Morales2014Renewables}, including scenario-based reserve requirements~\cite{Morales2009Reserves, Papavasiliou2011Reserve}.
Adaptive robust scheduling enforces the feasibility of balancing decisions over an uncertainty set~\cite{Jiang2012RobustUC,Bertsimas2013RobustSCUC}.
Chance-constrained formulations limit violations under partially known distributions~\cite{Lubin2016ChanceACOPF,Roald2015Weighted}.
For required $N$-$1$ contingencies, equilibrium feasibility typically remains a hard constraint.
Probabilities or risk measures address the remaining uncertainty.
Results remain sensitive to scenario construction, uncertainty-set calibration, and the treatment of capability uncertainty.

\subsubsection{Summary}

Static SCUC, SCED, and SCOPF form the baseline security-constrained scheduling models.
Current work primarily improves AC fidelity, contingency scalability, and uncertainty integration~\cite{Knueven2020UC,Aravena2023GO,Curtis2023SCACOPF}.
For IBRs, they schedule the operating point and any capability reserved for later response, but their equilibrium security claim does not establish dynamic performance.

\begin{table*}[t]
\caption{Literature summary of dynamic preventive operation}
\label{tab:dynamic_strands}
\centering
\scriptsize
\begin{tabularx}{\textwidth}{L{0.14\textwidth} L{0.17\textwidth} L{0.19\textwidth} L{0.22\textwidth} Y}
\toprule
\textbf{Topics} & \textbf{Security criteria} & \textbf{Scheduled decisions} & \textbf{Methods and literature} & \textbf{Strengths/Limitations} \\
\midrule
Frequency & RoCoF, nadir, quasi-steady-state; locational/multi-area response & Commitment, inertia/FFR enablement, headroom, BESS energy & Aggregate/multi-area models; MILP/MISOCP; data-driven nadir constraints~\cite{Anderson1990SFR,Teng2016FFR,Badesa2019FrequencyServices,Paturet2020LowInertiaUC,Tuo2023Locational,Shen2023DataDrivenNadir,Liu2024Nadir} & Direct scheduling link; delivery, saturation, duration, location \\
System strength and converter-driven stability & SCR/WSCR/gSCR; impedance; control-sensitive envelope & Synchronous/GFM commitment, topology, mode, reactive-current capability & Proxy-based UC/GFM allocation~\cite{CIGRE2016WeakGrid,Xin2017gSCR,NERCLowSCR2017,Kim2025StrengthUC,Cui2025GFM} & Locational link; depends on resource controls and the fault/EMT model \\
Small-signal stability & Spectral abscissa; damping ratio; modal margin & Dispatch, topology, stabilizer/IBR gains, control modes & Sensitivities; sequential optimization; surrogate-model stability limits~\cite{Rossi2024SSSCOPF,Agrawal2025SSS} & Modal-security link; mode switching, model changes, nonconvexity \\
Short-term voltage stability and recovery & Voltage dip/recovery; delayed recovery; ride-through & Reactive dispatch, commitment, IBR headroom, current-priority mode & Offline studies; analytical or surrogate-model limits; iterative time-domain assessment~\cite{NERC2022TVR,Kawabe2014STVSBoundary,Zhang2021STVSDeepLearning,Jiang2021STVSUC} & Trajectory criteria; load dynamics, saturation, model validity \\
Transient stability & Angle/energy margin; clearing time; trajectory & Dispatch, topology, protection settings, reserved emergency response & Direct discretization; simulation/cuts; reduced models; surrogate-model assessment~\cite{Gan2000TSCOPF,Pizano2010TSCOPF,Vu2016Lyapunov,Zhang2025TSCOPF,Wang2025DynamicsScheduling} & Dynamic criterion; scale, IBR phenomena, fidelity, validation \\
Multiple phenomena and offline-derived limits & Combined dynamic criteria; validated transfer, inertia, strength, or capability limits & Operating point, control mode, and enabled response & Multi-fidelity screening; operator studies; analytical or surrogate-model limits~\cite{Kundur2004Definition,Hatziargyriou2021Stability,NERC2025EMT,AEMO2025SecurityEnablement,Kawabe2014STVSBoundary,Zhang2021STVSDeepLearning} & Practical scheduling link; validity depends on configuration and validated operating range \\
\bottomrule
\end{tabularx}
\end{table*}

\subsection{Dynamic Preventive Operation}

Dynamic preventive scheduling selects operating conditions and resource configurations that satisfy prescribed post-disturbance performance criteria.
For IBRs, the schedule also preserves capability and configures the automatic response.
Table~\ref{tab:dynamic_strands} summarizes the representative work.

\subsubsection{Frequency-Security-Constrained Scheduling}

Frequency-security constraints are among the most established dynamic constraints embedded in scheduling because aggregate models link commitment and response to operational metrics~\cite{wang2024electric}.
A center-of-inertia frequency model typically has the form
\begin{equation}
\frac{2H_{\mathrm{sys}}}{f_0}\Delta\dot f(t)
=\Delta P_{\mathrm m}(t)+\Delta P_{\mathrm{FFR}}(t)
-\Delta P_{\mathrm{loss}}-D\Delta f(t),
\label{eq:swing_agg}
\end{equation}
where $\Delta f(t)$ is the system frequency deviation and $f_0$ is nominal frequency.
The parameters $H_{\mathrm{sys}}$ and $D$ are the aggregate inertia constant and the load-damping coefficient, respectively.
The terms $\Delta P_{\mathrm m}(t)$ and $\Delta P_{\mathrm{FFR}}(t)$ are the fast frequency responses, and $\Delta P_{\mathrm{loss}}>0$ is the generation loss.
Representative security constraints include rate of change of frequency (RoCoF), frequency nadir, and quasi-steady-state frequency deviation:
\begin{align}
|\Delta\dot f(0^+)|&\leq \mathrm{RoCoF}_{\max}, \\
\Delta f_{\mathrm{nadir}}&\ge -\Delta f_{\max}, \\
|\Delta f_{\mathrm{qss}}|&\leq \Delta f_{\mathrm{qss},\max}.
\end{align}
Here, $\Delta f_{\mathrm{nadir}}$ and $\Delta f_{\mathrm{qss}}$ are the nadir and quasi-steady-state frequency deviations.
The quantities $\mathrm{RoCoF}_{\max}$, $\Delta f_{\max}$, and $\Delta f_{\mathrm{qss},\max}$ are the maximum allowable RoCoF, nadir deviation, and quasi-steady-state deviation, respectively.

Early formulations used aggregate frequency-response models and minimum-inertia requirements~\cite{Anderson1990SFR}.
Later work jointly schedules inertia, primary frequency response, fast frequency response, headroom, and energy~\cite{Restrepo2005PrimaryRegulation,Teng2016FFR,Badesa2019FrequencyServices}.
Virtual-inertia and damping settings further expand this decision space~\cite{Poolla2017VirtualInertia,Paturet2020LowInertiaUC,Badesa2020Portfolio}.
Virtual inertia scheduling (VIS) illustrates how adjustable IBR settings and dispatch become preventive decisions~\cite{She2024VISRTED,She2025VISMicrogrid}.
The scheduled output and energy state determine the available headroom and response duration, subject to current and energy limits.
Recent work extends a scalar inertia requirement to coordinated response that accounts for timing and location.
These formulations approximate frequency nadir with data-driven models~\cite{Shen2023DataDrivenNadir,Liu2024Nadir} and include locational or multi-area effects~\cite{Tuo2023Locational,Wang2023MultiAreaFCUC}.

\subsubsection{System-Strength- and IBR-Driven-Stability-Constrained Scheduling}

System strength is inherently locational and control dependent.
Scheduling studies use several proxies for system strength.
These include fault level, short-circuit-ratio variants, equivalent grid impedance, and minimum commitment of synchronous or GFM resources~\cite{CIGRE2016WeakGrid,Xin2017gSCR,NERCLowSCR2017}.
For example, the formulation in~\cite{Kim2025StrengthUC} represents weighted SCR and network-impedance constraints within a MILP UC model.
Such proxies do not fully capture converter controls or interactions among equipment from different vendors~\cite{Rocabert2012ConverterControl,Rosso2021GFMReview}.
The same network condition may therefore be adequate for one IBR design but not another~\cite{Lin2020GFMRoadmap,ESIG2022GFM}.
Recent studies condition strength assessment on the operating point and IBR controls~\cite{Matevosyan2019GFM} and connect GFM allocation to operating cost~\cite{Cui2025GFM}.

Industrial practice is also evolving.
AEMO's Security Enablement Procedures document the operational enablement of security services in the National Electricity Market~\cite{AEMO2025SecurityEnablement}.
Its inertia and system-strength documents provide the related requirements and guidance~\cite{AEMO2023Strength,AEMO2024Inertia,AEMO2026SecurityGuidelines}.

\subsubsection{Small-Signal-Stability-Constrained Scheduling}

Dynamic security can be tied directly to a linearized system model.
Small-signal stability is commonly expressed through the state matrix $\mathbf A(\boldsymbol{\vartheta})$ obtained at the scheduled equilibrium:
\begin{equation}
\Delta\dot{\mathbf x}=\mathbf A(\boldsymbol{\vartheta})\Delta\mathbf x,
\end{equation}
where $\Delta\mathbf x$ is the state-variable deviation vector.
The corresponding constraints on eigenvalue real parts and modal damping ratios are
\begin{equation}
\max_k \operatorname{Re}\{\lambda_k(\mathbf A)\}\leq -\epsilon,
\qquad
\zeta_k\geq\zeta_{\min}.
\label{eq:sss}
\end{equation}
Here, $\lambda_k$ and $\zeta_k$ are the eigenvalue and damping ratio of mode $k$, respectively.
The constants $\epsilon>0$ specifies the required eigenvalue margin, and $\zeta_{\min}$ specifies the minimum damping ratio.
Methods include eigenvalue sensitivities, sequential optimization, and surrogate models of the stability boundary.
Regression-based OPF models can represent a learned stability boundary as a continuous nonlinear constraint without adding a large set of binary variables~\cite{Rossi2024SSSCOPF}.
More recent work explores convex polynomial approximations and multistage learning~\cite{Agrawal2025SSS}.

Critical modes can appear, disappear, or exchange identity as topology, dispatch, and controller settings change.
The feasible set is nonconvex and may be nonsmooth at mode switching.
Model parameters for IBR controls may be unavailable or vendor confidential.
A small-signal model also does not establish large-disturbance or fault-trajectory performance.
IBR parameters should be treated as scheduling decisions only when they are operationally adjustable within a validated operating range.

\subsubsection{Short-Term Voltage-Stability-Constrained Scheduling}

Dynamic preventive formulations constrain post-disturbance voltage trajectories, including transient voltage dips, recovery time, and ride-through performance~\cite{VanCutsem1998Voltage, Hatziargyriou2021Stability}.
IBR reactive current injection, current saturation, and relay logic shape these trajectories.
Direct embedding of detailed voltage dynamics in UC remains limited.
Offline studies and analytical boundaries provide operating limits for scheduling~\cite{NERC2022TVR,Kawabe2014STVSBoundary}.
Other work uses surrogate-model classifiers or iterates between scheduling and time-domain simulation~\cite{Zhang2021STVSDeepLearning,Jiang2021STVSUC}.
Available reactive-current response depends on the scheduled $P$--$Q$ operating point and current-priority logic.
Treating active- and reactive-current support independently can overstate capability.
Hence, direct integration into scheduling remains application-specific.

\subsubsection{Transient-Stability-Constrained Scheduling}

Transient-stability constraints are required when large disturbance trajectories cannot be represented by linearization criteria.
As summarized in~\cite{Zhang2025TSCOPF}, transient-stability-constrained optimal power flow (TSCOPF) traditionally uses rotor-angle separation, critical clearing time, or transient energy.
Direct discretization converts DAEs into a large nonlinear program.
Sequential methods alternate OPF and time-domain simulation and add sensitivity-based constraints.
Other methods use simulation-based search and machine-learning approximations of stability margins and feasible boundaries.
Their foundations include transient simulation and energy-function models~\cite{Dommel1972Transient,Bergen1981Structure,Pavella2000Transient}.
Other work uses direct stability-constrained OPF~\cite{Gan2000TSCOPF,Chiang2011Direct} or trajectory- and Lyapunov-based assessment~\cite{Pizano2010TSCOPF,Vu2016Lyapunov}. They provide a direct link between one operating point and the post-disturbance stability.

Transient-stability-constrained unit commitment (TSCUC) compounds the dynamic problem with multiperiod binary commitment.
Practical approaches therefore use reduced-order constraints or selected contingencies.
The dynamics-incorporated scheduling framework in~\cite{Wang2025DynamicsScheduling} separates reusable dynamic simulation from the scheduling master and returns enforceable stability constraints.
This solution structure is more scalable than duplicating full dynamic states for every commitment interval and contingency.

For IBR-dominated systems, rotor-angle metrics alone are insufficient when instability is driven by phase-locked-loop (PLL) loss of synchronism, dc-link dynamics, or control saturation.
Electromagnetic-transient (EMT) models may be needed for weak-grid phenomena and fast converter-control interactions.
A key computational issue is how to incorporate EMT-based assessment into scheduling without embedding full EMT models.
TSCOPF is more mature than TSCUC, while EMT simulation is typically used for external validation.

\subsubsection{Summary}

Aggregate frequency-security constraints are the most established dynamic constraints used directly in scheduling.
Other IBR-driven phenomena more often use reduced-order constraints or external RMS or EMT assessment within a validated operating range.
IBR capability depends on the scheduled operating point, shared device limits, and enabled response.
Preserving the response may restrict dispatch through these limits.
The reviewed studies seldom assess model validity and response delivery together.

\subsection{Solution Frameworks for Preventive Scheduling}

\begin{figure}[t]
\centering
\includegraphics{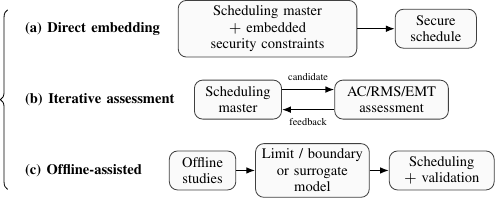}
\caption{Solution frameworks for coupling preventive scheduling with security assessment.}
\label{fig:preventive_solution_frameworks}
\end{figure}

Preventive solution frameworks determine which security constraints remain in the scheduler and which are evaluated externally.
For IBRs, scheduling and security assessment must use consistent operating points, controller configurations, and capability limits.
Fig.~\ref{fig:preventive_solution_frameworks} summarizes three structures.

\subsubsection{Direct Embedding}

Direct embedding places algebraic, reduced-order, or discretized trajectory constraints inside the scheduling problem.
It gives the optimizer direct access to the security limits and is effective when the selected constraints remain compact.
Its use becomes difficult as dynamic states, periods, and contingencies accumulate~\cite{Knueven2020UC, Capitanescu2016ACOPF}.

\subsubsection{Iterative Security Assessment}

An iterative framework separates the scheduling master from AC, RMS, or EMT assessment.
The master proposes an operating point and enabled capability, while security subproblems return violations, sensitivities, or margins.
This structure supports higher-fidelity assessment, but cut validity may be lost after a change in commitment, topology, or IBR control mode~\cite{Phan2014EfficientSCOPF,Velloso2021Decomposition,Jiang2021STVSUC}.

\begin{table*}[t]
\caption{Representative formulations for static corrective operation}
\label{tab:static_corrective}
\centering
\scriptsize
\begin{tabularx}{\textwidth}{L{0.15\textwidth} L{0.17\textwidth} L{0.19\textwidth} L{0.21\textwidth} Y}
\toprule
\textbf{Formulation emphasis} & \textbf{Security criteria} & \textbf{Corrective decisions} & \textbf{Methods and literature} & \textbf{Strengths/Limitations} \\
\midrule
Continuous corrective actions & Post-contingency AC/DC equilibrium; balance, thermal, voltage, action limits & Redispatch, controllable $P$--$Q$ setpoints, storage response, and curtailment & Corrective SCOPF/SCUC; sparse actions~\cite{Monticelli1987PCSCOPF,Capitanescu2008Corrective,Fu2006ContingencyDispatch,Chen2017CorrectiveRecourse,Wen2016StorageSCUC,Phan2015SparseCorrective,Alizadeh2022SecurityControl} & Coordinates modeled actions; activation, trajectory, sustainment not established \\
Mixed continuous--discrete corrective actions & Equilibrium under selectable configurations & Redispatch/setpoints; switching and discrete settings & Security-control and corrective-switching models~\cite{Li2017CorrectiveSwitching,Alizadeh2022SecurityControl} & Broader modeled action set; configuration-dependent feasibility \\
Preventive--corrective coordination & Base-state and contingency equilibrium feasibility & Base schedule/configuration; contingency-specific corrective targets & Preventive--corrective SCOPF/SCUC~\cite{Monticelli1987PCSCOPF,Fu2006ContingencyDispatch,Chen2017CorrectiveRecourse,Xu2014PCSCOPF,Xu2016Partitioning,Thomas2015FlexibleSCOPF,Alizadeh2023Tractable} & Cost--modeled-capability tradeoff; equilibrium-only claim \\
\bottomrule
\end{tabularx}
\end{table*}

\subsubsection{Offline Approximation and Online Validation}

Offline studies can produce operating limits, security boundaries, or surrogate models that are enforced in online scheduling.
An independent higher-fidelity assessment checks the resulting schedule.
Validity depends on approximation error and on whether the topology, controller configuration, and operating point remain within the validated range~\cite{NERC2022TVR,Kawabe2014STVSBoundary,Zhang2021STVSDeepLearning,Popli2024MLProxies}.

\section{Security-Constrained Corrective Operation}
\label{sec:corrective}

This section examines static and dynamic corrective operation, followed by solution frameworks for preparing and implementing corrective actions within their application timelines.

\subsection{Static Corrective Operation}

Static corrective operation uses contingency-specific actions to establish an acceptable post-contingency equilibrium.
Table~\ref{tab:static_corrective} summarizes representative formulations and their strengths and limitations.

\subsubsection{Static Corrective Formulations}

Static corrective formulations include actions selected after the contingency is identified.
In optimization, the associated contingency-specific variables are often called recourse variables.
This paper refers to them as corrective action variables.
They may represent changes in power injection, network configuration, or controllable setpoints.
Each resulting equilibrium must satisfy thermal, voltage, and action limits.

Corrective actions can be included in SCUC, SCED, or SCOPF.
Commitment is fixed before the contingency, while modeled actions include redispatch, load response, and network switching~\cite{Fu2006ContingencyDispatch, Chen2017CorrectiveRecourse, Wen2016StorageSCUC}.
Generation redispatch is the conventional continuous corrective action~\cite{Monticelli1987PCSCOPF}.
The action set can include active- and reactive-power setpoint changes, renewable curtailment, and storage injection~\cite{Capitanescu2008Corrective,Phan2015SparseCorrective,Wen2016StorageSCUC,Alizadeh2022SecurityControl}.
For IBRs, energy, control mode, and response time determine whether a target can be reached and sustained.
Switching and other discrete settings broaden the action set but increase computational burden because feasibility depends on the selected configuration~\cite{Li2017CorrectiveSwitching, Alizadeh2022SecurityControl}.
Continuous and discrete actions can be combined within one contingency model~\cite{Alizadeh2022SecurityControl,Alizadeh2023Tractable}.
A contingency-indexed action may be precomputed.
It remains static corrective when only the resulting equilibrium is assessed.

\subsubsection{Preventive--Corrective Coordination}

The base operating point and scheduled configuration determine the corrective actions available after each contingency.
Preventive--corrective SCOPF jointly selects the base schedule and contingency-specific corrective targets, balancing preventive cost against modeled corrective capability~\cite{Monticelli1987PCSCOPF,Capitanescu2008Corrective,Xu2014PCSCOPF}.
Its security claim is limited to the modeled post-contingency equilibrium.
Operational use also requires the resource to reach the target within the required time.

Later studies improve scalability through contingency partitioning~\cite{Xu2016Partitioning} and approximations for larger models~\cite{Alizadeh2023Tractable}.
Other extensions allow flexible or multiperiod corrective actions~\cite{Thomas2015FlexibleSCOPF, Alizadeh2022SecurityControl}.
Risk-based variants use contingency probabilities to weight corrective costs and consequences~\cite{Wen2014Discussion,Xu2014Closure}.
The required post-contingency criteria remain enforced independently of this probability weighting.

\subsubsection{Summary}

Static corrective formulations assess post-contingency equilibria under continuous or discrete corrective actions.
For IBRs, these actions remain within the capability preserved by the preventive schedule.
Equilibrium feasibility does not establish a secure transition.
This limitation motivates the trajectory-oriented corrective strategies.

\subsection{Dynamic Corrective Operation}

Dynamic corrective operation uses the measured post-contingency state to select an action beyond the preconfigured automatic response.
For an IBR, this decision may update a power reference or a controller setting.
Table~\ref{tab:dynamic_corrective} summarizes the reviewed literature by security criteria, decisions, methods, and limitations.

\begin{table*}[t]
\caption{Literature summary of dynamic corrective operation}
\label{tab:dynamic_corrective}
\centering
\scriptsize
\begin{tabularx}{\textwidth}{L{0.17\textwidth} L{0.15\textwidth} L{0.19\textwidth} L{0.21\textwidth} Y}
\toprule
\textbf{Topics} & \textbf{Security criteria} & \textbf{Corrective decisions} & \textbf{Methods and literature} & \textbf{Strengths/Limitations} \\
\midrule
Post-event supervisory action selection & Trajectory-based stability or selected dynamic limits & Selects an action beyond fixed logic & Supervisory emergency-control selection~\cite{Wehenkel2004PreventiveEmergency,Genc2010PreventiveCorrective} & Fast response; limited coverage and action-feasibility checks \\
Feedback emergency control & Frequency/voltage trajectories; transient or damping limits & Measurements update commands or controller settings & Emergency and voltage control; fast power modulation~\cite{Vittal1989Emergency,Kundur1994,VanCutsem1998Voltage,Glavic2011MPC,Martin2017CorrectiveMPC} & Direct trajectory control; saturation, model error, communication, backup response \\
Online optimization and MPC & Frequency, voltage, or post-fault trajectories; controller and terminal limits & State estimate updates IBR/storage references and other controls & System- and converter-level MPC~\cite{Almassalkhi2015CascadeMPC,Martin2017CorrectiveMPC,AdemolaIdowu2021MPC,Stanojev2022MPC,ArjomandiNezhad2024GFM} & Coordinated control; solution time, estimation error, model validity, infeasibility \\
Learning-assisted corrective policy & Trajectory or terminal security within a validated range & Measurements select an action or approximate a policy & Neural Lyapunov corrective control~\cite{Bellizio2023NeuralLyapunov} & Fast evaluation; false-secure cases, validated-range limits, and approval \\
\bottomrule
\end{tabularx}
\end{table*}

\subsubsection{Post-Event Action Selection and the RAS Boundary}

RAS logic designed and armed in advance is a preventive configuration.
Response logic fixed before the event remains a preconfigured automatic response, even when measured conditions activate different predefined branches~\cite{Wehenkel2004PreventiveEmergency,Genc2010PreventiveCorrective}.
A corrective decision occurs when online optimization selects an action beyond that preconfigured logic.
RASs can trip or run back generation, shed load, and reconfigure the network~\cite{CIGRE2001SPS,Zima2005SIPS,Andersson2005Blackout,NERC2020RAS}.
NERC requirements establish design assessment and periodic review for approved schemes~\cite{NERCPRC012}.
These sources document requirements for preconfigured schemes.

\subsubsection{Emergency Optimization and Feedback Control}

Feedback and online optimization adapt corrective action to the observed trajectory.
Classical emergency control uses transient simulation to select actions.
Energy functions and Lyapunov methods provide related assessment tools~\cite{Dommel1972Transient,Bergen1981Structure,Vittal1989Emergency,Pavella2000Transient,Chiang2011Direct,Vu2016Lyapunov}.
Voltage-control methods coordinate network controls and load relief with generation or HVDC response~\cite{VanCutsem1998Voltage,Glavic2011MPC,Martin2017CorrectiveMPC}.
At the system level, model predictive control (MPC) coordinates generation, storage, and controllable load to mitigate thermal-overload cascades~\cite{Almassalkhi2015CascadeMPC}.
For frequency control, centralized, decentralized, and explicit MPC formulations can update the active-power references of constrained IBRs~\cite{AdemolaIdowu2021MPC,Stanojev2022MPC}.
At the converter level, post-fault MPC can adjust phase angle and active-power references when current saturation threatens GFM transient recovery~\cite{ArjomandiNezhad2024GFM}.
Available energy and saturation bound the IBR capability available to these controllers.
MPC generalizes these actions by repeatedly solving a finite-horizon problem
\begin{align}
\min_{\{\mathbf u_k\}_{k=0}^{N-1}} \quad & \sum_{k=0}^{N-1}
\ell(\mathbf x_k,\mathbf u_k)+V_{\mathrm f}(\mathbf x_N) \\
\text{s.t.}\quad & \mathbf x_{k+1}=\mathbf f_{\mathrm d}(\mathbf x_k,\mathbf u_k),\quad
\mathbf x_k\in\mathcal X,\quad \mathbf u_k\in\mathcal U, \nonumber\\
& k=0,\ldots,N-1,\qquad \mathbf x_N\in\mathcal X_{\mathrm f},
\end{align}
and implementing the first control move.
Here, $\mathbf x_k$, $\mathbf u_k$, and $\ell$ are the predicted state, control input, and stage cost, respectively.
The state-update equation uses the discrete-time prediction model $\mathbf f_{\mathrm d}$, while $\mathcal X$ and $\mathcal U$ are the state and control-input constraint sets.
The symbols $N$, $V_{\mathrm f}$, and $\mathcal X_{\mathrm f}$ denote the prediction horizon, terminal cost, and terminal set, respectively.
MPC coordinates multivariable actions under controller and terminal constraints, but requires accurate state estimation and a solution within the required response time.

\subsubsection{Learning-Assisted Corrective Control}

Learning-assisted methods can classify the state, approximate a policy, or provide a stability certificate~\cite{she2022fusion}.
Convolutional neural networks have been used to identify transient-stability status and instability modes from post-disturbance measurements~\cite{Shi2020CNNTransientStability}.
Safe reinforcement learning has also been applied to emergency load shedding for post-fault voltage recovery~\cite{Vu2021SafeRLLoadShedding}.
In addition, neural Lyapunov methods embed a learned certificate in corrective optimization~\cite{Bellizio2023NeuralLyapunov}.
The certificate defines a trajectory condition within a state region verified offline.
Its security claim is therefore limited to that range, so validation should report false-secure cases and out-of-range performance.

\subsubsection{Summary}

Preconfigured RASs have established design and assessment procedures, and post-event control is generally less investigated.
The reviewed methods use the identified contingency and measured state to update emergency actions, but their evidence remains application-specific.
Feedback control acts on measured trajectories, while MPC predicts the trajectory and enforces controller limits online.
Offline security regions and learned certificates can reduce the online assessment burden, but their security claim remains limited to the validated operating range~\cite{Genc2010PreventiveCorrective, Xie2020CoordinatedDispatch, Bellizio2023NeuralLyapunov}.
For IBRs, the feasible corrective action depends on the capability remaining after automatic response.
Accordingly, a dynamic corrective formulation should link the preventive schedule and post-contingency trajectory to the corrective capability available after the event.

\subsection{Solution Frameworks for Corrective Operation}

Corrective solution frameworks differ primarily in whether the action is precomputed or optimized after the contingency.
Hybrid implementations combine offline design with online update.
In all cases, the action must be feasible and available within the required response time.

\subsubsection{Precomputed Actions}

Precomputed methods prepare contingency-specific actions for a defined set of contingencies and operating conditions.
Static corrective actions can be evaluated through decomposition, contingency partitioning, and sparse formulations~\cite{Capitanescu2008Corrective,Xu2014PCSCOPF,Xu2016Partitioning,Phan2015SparseCorrective}.
Dynamic actions can likewise be prepared and assessed offline.
Preconfigured response logic remains automatic, whereas post-event selection or modification is corrective~\cite{Zima2005SIPS,NERC2020RAS}.
Validity is limited to the contingencies, operating conditions, and controller configurations covered by the offline assessment.

\subsubsection{Online Corrective Optimization}

Online corrective optimization uses the identified contingency and post-contingency state to calculate the corrective action.
Online static corrective optimization determines a post-contingency equilibrium, whereas online dynamic corrective optimization also represents the predicted trajectory and controller limits.
Linearized models and contingency-specific control selection reduce the size of the online MPC problem~\cite{Glavic2011MPC, Martin2017CorrectiveMPC}.
A learned stability certificate can move part of the assessment offline and leave a smaller corrective optimization online~\cite{Bellizio2023NeuralLyapunov}.
The feasible action set must be updated from the actual IBR operating point and the capability already used by the automatic response.
Online optimization increases adaptability but places stronger requirements on state estimation, model accuracy, and computation time.

The response-time requirement covers each phase of the corrective operation~\cite{Zima2005SIPS,Martin2017CorrectiveMPC}.
A practical framework must detect infeasible cases and specify a fallback when the intended action is unavailable~\cite{NERCPRC012}.

\section{IBR Capability Effects and Research Directions}
\label{sec:closed_loop}

This section discusses IBR capability effects and identifies the challenges and critical research needs.

\subsection{IBR Capability Effects}

\begin{figure}[t]
\centering
\includegraphics[width=\columnwidth]{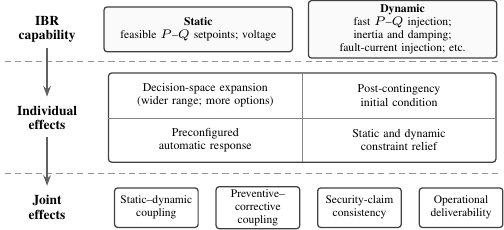}
\caption{IBR capability and its effects on security-constrained operation.}
\label{fig:cross_quadrant_synthesis}
\end{figure}

IBR capability is conditional on the operating point and controller configuration.
Static capability comprises feasible steady-state $P$--$Q$ and voltage setpoints for power flow control.
Dynamic capability includes fast $P$--$Q$ injection, inertia and damping support, fault-current injection, etc.~\cite{Lin2020GFMRoadmap, IEEE2800_2022, she2026review}.
These capabilities affect individual formulations and their coordination, as summarized in Fig.~\ref{fig:cross_quadrant_synthesis}.

\subsubsection{Effects Within Individual Formulations}

IBR capability affects the formulation through four channels.
First, it expands the decision space.
This expansion can widen the design and preventive-configuration range in $\mathcal F_0$ or add corrective options to $\mathcal A_c$.
Second, it changes the post-contingency initial condition through $\Gamma_c$.
Third, it changes the preconfigured automatic response through the equilibrium model or $\boldsymbol\mu_c$.
Fourth, it can relieve a constraint in $\mathcal S_c^{\mathrm{stat}}$ or $\mathcal S_c^{\mathrm{dyn}}$.
Within an individual formulation, crediting these capabilities can lower operating cost or improve security performance relative to a conventional formulation that omits them.

\subsubsection{Joint Effects Across Formulations}

Joint effects arise when several formulations rely on the same IBR capability.
Shared power and energy limits couple security requirements and decision stages~\cite{She2024VISRTED, She2025VISMicrogrid, Cui2025ControlMode}.
Use of capacity at equilibrium can reduce the response available for dynamic security.
Conversely, preserving dynamic response can restrict the static operating point.
Across decision stages, the preventive schedule determines the automatic response and the corrective capability that remains after the contingency.
Anticipated corrective capability can reduce the preventive margin required to satisfy a security constraint.
The net effect is therefore not necessarily an expansion of the feasible operating region of joint formulations.
These couplings can make results that are feasible in isolation infeasible when the formulations are combined~\cite{Capitanescu2007Viability,Capitanescu2009DynamicSimulation}.

Therefore, joint scheduling should establish that the credited response is operationally deliverable.
This requires a consistent operating condition, response assumption, and set of device limits across formulations.
The credited response must remain available after the automatic response and be executable within the required time~\cite{IEEE2800_2022, Zima2005SIPS, Martin2017CorrectiveMPC}.

\subsubsection{Summary}
IBR capability affects both individual security-constrained formulations and coupled formulations.
Within an individual formulation, IBR capability can expand the feasible decision set or relieve a security constraint.
Across coupled formulations, this benefit is retained only when the credited response remains available under shared device limits and can be delivered within the required time.
The first level explains how capability changes a security-constrained formulation.
The second determines whether the modeled benefit is operationally deliverable.

\subsection{Research Directions}

Three critical research needs for integrating IBR capability into security-constrained operation are discussed below.

\subsubsection{IBR Capability Characterization and Quantification}
\leavevmode\par
\textbf{\textit{Challenge:}}
IBR control development and system-level security-constrained operation are often treated separately.
Even when a grid-support capability has been demonstrated, its availability across operating points and controller configurations are not consistently quantified under shared device limits~\cite{Badesa2020Pricing}.
Without this quantified capability, potentially useful responses cannot be consistently scheduled in preventive and corrective operation.

\textbf{\textit{Vision:}}
Develop IBR control algorithms that provide grid-support responses targeted to binding system-security constraints.
Quantify each response across operating points and controller configurations.
Then, translate the quantified capability into security constraints or feasible action sets for system-level operation models.
This process would enable demonstrated IBR capability to be quantified, integrated into operation, and appropriately valued in electricity markets.

\subsubsection{Joint Scheduling Across Formulations}
\leavevmode\par
\textbf{\textit{Challenge:}}
Separate security-constrained formulations can allocate the same device capability to incompatible functions.
They can also assume different operating conditions or device limits.
Differences in information and response time can further leave the intended action infeasible or unavailable.

\textbf{\textit{Vision:}}
Use consistent operating conditions and shared device limits across formulations.
Coordinate the pre-contingency configuration with automatic and corrective responses~\cite{Wehenkel2004PreventiveEmergency,Genc2010PreventiveCorrective,Xu2014PCSCOPF,Martin2017CorrectiveMPC}.
State the information and response time available at each stage.
Specify a fallback when the intended action is unavailable or late~\cite{NERCPRC012}.
End-to-end studies should verify the sequence from event detection through actuation.

\subsubsection{Scalable and Verifiable Solution Frameworks}
\leavevmode\par
\textbf{\textit{Challenge:}}
Multi-timescale scheduling can become too large when each contingency includes a dynamic trajectory and corrective policy.
A high-fidelity model is generally too large to solve directly.
Scenario reduction and model approximation improve scalability but may miss a binding security constraint~\cite{Aravena2023GO,Wang2025DynamicsScheduling,Martin2017CorrectiveMPC,Popli2024MLProxies}.

\textbf{\textit{Vision:}}
Develop multi-timescale methods that retain the variables linking scheduling and security assessment.
Use higher-fidelity assessment when it can affect the scheduling decision.
Co-simulation across device, EMT, and scheduling models can provide consistent interfaces for validating corrective actions while preserving computational tractability~\cite{wang2021transmission}.
In addition, offline policy preparation can be combined with online selection or adjustment.
AI methods may support security screening, evaluation of surrogate models, and selection of corrective policies when operating and security constraints are enforced explicitly~\cite{oboreh2023virtual, Donti2021DC3, Popli2024MLProxies}.
Cases near the security boundary still require applicability checks and higher-fidelity validation.
Benchmark studies should report model versions, validated operating ranges, and contingency coverage~\cite{NERC2025EMT, NERC2025ModelQuality, Baba2019PGLib, Elbert2024GODataset}.

\section{Conclusion}
\label{sec:conclusion}

Security-constrained operation is evolving to include IBR capability and dynamic security.
This paper presents a two-axis view: static versus dynamic security and preventive versus corrective decisions.
A generic formulation links the four categories and shows how IBR capability contributes to each.
The synthesis distinguishes capability effects within individual formulations from joint effects across coupled formulations.
Within one formulation, an IBR response may expand the decision space or relieve a security constraint.
Across formulations, shared device limits and response assumptions can make results that are feasible in isolation infeasible when the formulations are coupled.

New IBR capability development should be accompanied by assessment across security requirements and decision stages.
Operational credit requires validation for the modeled condition and a response that remains deliverable after the contingency.
The resulting research needs concern capability characterization and quantification, joint scheduling, and scalable solution frameworks.

\bibliographystyle{IEEEtran}
\bibliography{references}

\end{document}